\documentclass[graybox,natbib,journals]{svmult} 
\usepackage{graphicx}
\usepackage{floatrow}
\usepackage{amssymb}

\usepackage{makeidx}         
\usepackage{graphicx}        
\usepackage{multicol}        
\usepackage[bottom]{footmisc}

\makeindex             

\gdef\sb{mag\,arcsec$^{-2}$}

\newcommand\aap{Astronomy \& Astrophysics }

\newcommand\apjl{ApJ }
\newcommand\apjs{ApJSupp }

\begin{document}

\title*{Future Prospects: Deep Imaging of Galaxy Outskirts using Telescopes Large and Small}
\titlerunning{Deep Imaging of Galaxy Outskirts using Telescopes Large and Small}
\authorrunning{Team Dragonfly}

\author{Roberto Abraham$^{1,2}$, Pieter van Dokkum$^3$, Charlie Conroy$^4$,
Allison Merritt$^3$, Jielai Zhang$^{1,2,5}$, Deborah Lokhorst$^{1,2}$, Shany Danieli$^{3,4,6,7}$, 
Lamiya Mowla$^3$}
\institute{$^1$Department of Astronomy and Astrophysics, University of Toronto, 50 St. George Street, Toronto, ON M5S 3H4, Canada.\\ Primary contact: abraham@astro.utoronto.ca \\
$^2$Dunlap Institute for Astronomy and Astrophysics, University of Toronto, 50 St. George Street, Toronto, ON M5S 3H4, Canada\\
$^3$Department of Astronomy, Yale University, 
Steinbach Hall, 52 Hillhouse Avenue, New Haven, CT 06511, USA\\
$^4$Harvard-Smithsonian Center for Astrophysics, 60 Garden St, Cambridge, MA 02138, USA\\
$^5$Canadian Institute for Theoretical Astrophysics, 60 St. George Street, Toronto, ON M5S 3H8, Canada\\
$^6$Department of Physics, Yale University, 266 Whitney Avenue New Haven, CT 06511, USA\\
$^7$Yale Center for Astronomy and Astrophysics, 52 Hillhouse Avenue, New Haven, CT 06511, USA\\
}

\maketitle

\abstract{
The Universe is almost totally unexplored at low surface brightness levels. In spite of great progress in the construction of large telescopes and improvements in the sensitivity of detectors, the limiting surface brightness of imaging observations has remained static for about forty years. Recent technical advances have at last begun to erode the barriers preventing progress. In this Chapter we describe the technical challenges to low surface brightness imaging, describe some solutions, and highlight some relevant observations that have been undertaken recently with both large and small telescopes.  Our main focus will be on discoveries made with the Dragonfly Telephoto Array (Dragonfly), which is a new telescope concept designed to probe the Universe down to hitherto unprecedented low surface brightness levels. 
We conclude by arguing that these discoveries are probably only scratching the surface of interesting phenomena that are observable when the Universe is explored at low surface brightness levels. 
}

\section{Motivation}

A fundamental prediction of hierarchical galaxy formation models in a dark energy-dominated Cold Dark Matter cosmology ($\rm \Lambda$CDM) is that all galaxies are surrounded by a vast and complex network of ultra-low surface brightness filaments and streams, the relics of past merger events (\citealt{purcell:07, johnston:08, cooper:13, pillepich:14}). Resolved (star count-based) analyses have revealed the existence of streams and filaments around the Milky Way (e.g., \citealt{majewski:03, belokurov:07, carollo:07, bell:08}) and M31 (e.g., \citealt{ibata:01, ferguson:02, ibata:07, richardson:08, mcconnachie:09, gilbert:12}). These impressive results provide strong evidence that some massive spiral galaxies formed, at least in part, hierarchically. In the $\rm \Lambda$CDM picture many thousands of such streams and filaments combine over time to define galactic extended stellar haloes, with the bulk of the material distributed at surface brightnesses well below 30\,mag/arcsec$^2$ (\citealt{johnston:08}). In contrast to the star count-based results on stellar streams, the detection of these haloes has proven elusive. The {\it Hubble Space Telescope} ({\it HST}) has undertaken some very successful deep pencil beam star count surveys of a number of galaxies outside the Local Group (e.g., the GHOSTS survey; c.f. 
\citealt{radburn:11} and references therein) but since these stellar haloes are very extended (tens of arcminutes for nearby objects), nearby haloes may not be well sampled by pencil beams, and the seemingly much simpler strategy of trying to image them directly using ground-based telescopes is quite attractive. Many such studies have claimed detections of the extended low surface brightness stellar haloes of galaxies on the basis of direct imaging, but these claims are now controversial, with recent investigations dismissing these ``haloes'' as simply being scattered light. This point was first made by de Jong in 2008, and the putative detection of low surface brightness stellar haloes from unresolved imaging has recently been the subject of two exhaustive investigations by 
\citet{sandin:14,sandin:15}, who concludes that most claimed detections are spurious.

\section{Why is Low Surface Brightness Imaging Hard?}

The faintest galaxies detected at present are about seven magnitudes (over a factor of 600) fainter than the faintest galaxies studied during the ``photographic era'' of astronomy prior to the mid-1980s. It is therefore quite remarkable that over this same period of time there has been essentially no improvement in the limiting surface brightness of deep imaging observations of galaxies. (For example, the low surface brightness limits presented in 
\citealt{kormendy:74} are quite impressive even by modern standards). As we will describe below, this is because low surface brightness imaging is not usually limited by photon statistics or by spatial resolution. Instead, it is limited by imperfect control of systematic errors. The implication is that consderable work can be done by relatively small telescopes, and indeed a number of interesting investigations of tidal structures around nearby galaxies have appeared recently, based on amateur-professional collaborations using small telescopes (\citealt{martinez-delgado:09, martinez-delgado:10, martinez-delgado:12, martinez-delgado:16}). 

\begin{figure}[tbh]
\begin{center}
{\vspace*{0.2cm}\includegraphics[width=4.6in]{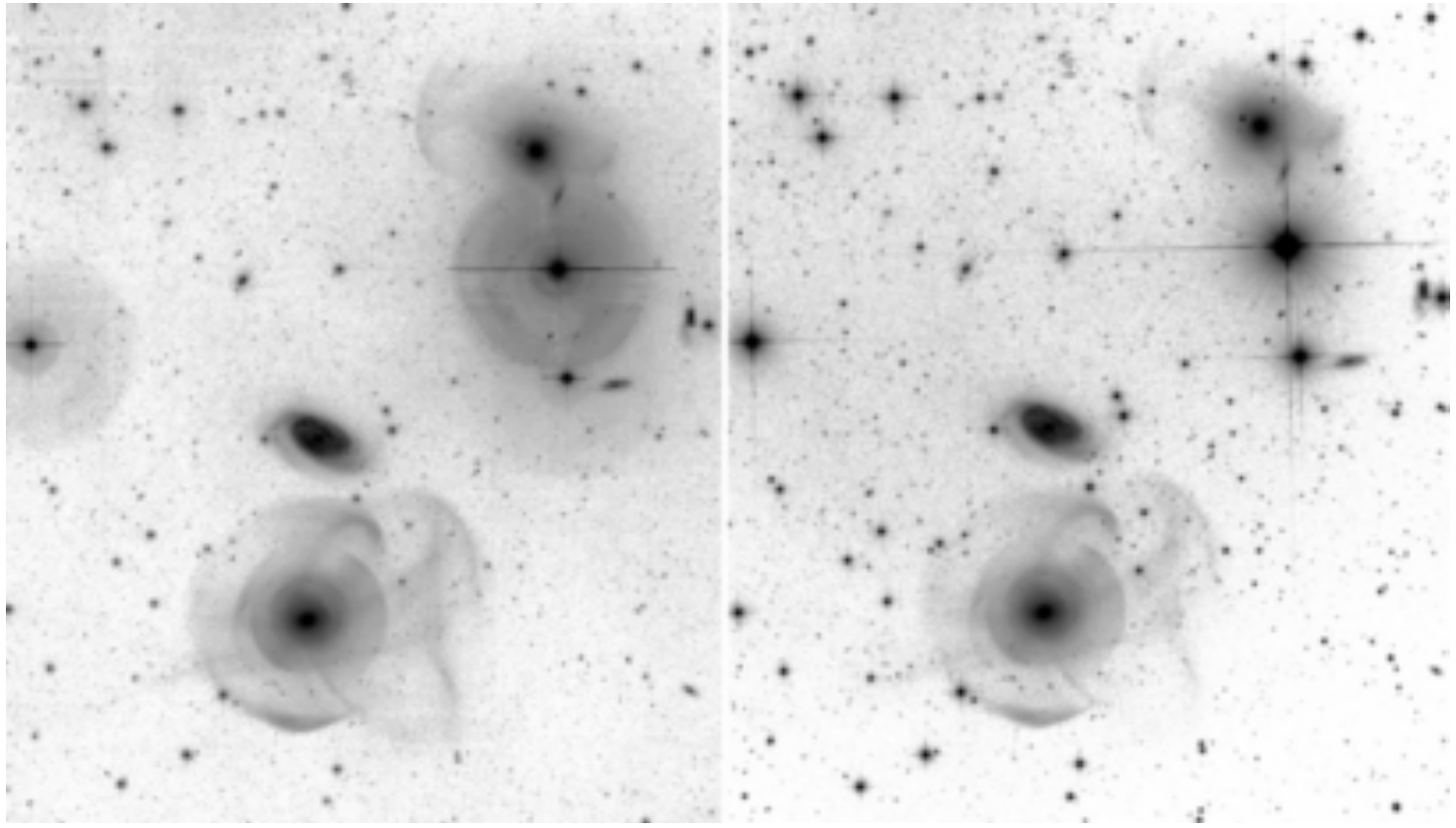}}
{\caption{
Images of the same field around NGC~474 (to the East) and NGC~467 (to the West) obtained with two different cameras and telescopes: MegaCam on the 3.6\,m CFHT ({\it left}) and a 12\,inch amateur telescope ({\it right}). The integration time is $30\times$ longer on the small telescope than on the CFHT, but the limiting surface brightness is the same, and note the absence of large stellar haloes in the image obtained by the amateur telescope which confuse the CFHT data at low surface brightness levels. Figure reproduced with permission from Duc et al. (2015)
}}
\end{center}
\label{fig:halos}
\end{figure}

A major benefit of the work done to date using small telescopes is that the imaging systems tend to be relatively simple. The absence of complicated re-imaging optics\footnote{Re-imaging optics are an essential component of  cameras on large telescopes for many reasons, e.g., to reduce their long native focal lengths in order to provide reasonable image scales on relatively small sensors. Most small telescopes utilize small and simple flat field correctors, or even image straight onto their sensors, since some small telescope optical designs have intrinsically well-corrected flat focal planes that are a good match to CCD pixel sizes.}  tends to produce images from small telescopes that are ``cleaner'' than those of large telescopes, with fewer internal reflections. (See, for example, the very striking examples in \citealt{duc:15}, one of which is reproduced in Fig.~1). As will be described below, this clean focal plane is an important benefit, because internal reflections and scattering are important systematics that have played the major role in defining the limiting surface brightness of all visible-wavelength imaging for the past few decades (\citealt{slater:09}). For certain classes of observation (such as the detection of intra-cluster light, or of stellar haloes) the impact of ghosting and scattering on the wide-angle point spread function\index{point spread function} (PSF) can have a huge impact on the credibility of the results. 

On the other hand, provided one is content with probing the Universe at surface brightness levels brighter than $V\sim28.5$\,\sb~(which has effectively been the limit for the last 40 years), ghosting and internal reflections are generally an inconvenience rather than a show-stopper. The well-motivated astronomer can take steps to limit the impact of ghosting on large telescopes by using optimal dithering strategies (\citealt{tal:11}), and in extreme cases one can attempt to model internal reflections and subtract them off (\citealt{slater:09, duc:16, roman:16}). Even when such steps are not taken, in general it is fairly clear when a compact structure on an image is due to scattered light or an internal reflection, as opposed to something intrinsic to a galaxy, such as a tidal feature.
All things considered, the authors of this chapter would certainly rather base conclusions about the prevalence of most galactic features (with the very notable exception of stellar haloes) on large and fair samples of low-intermediate redshift galaxies that can be obtained efficiently by large telescopes such as the Canada-France-Hawaii Telescope (CFHT; e.g., \citealt{atkinson:13}), rather than base conclusions on ``cleaner'' images obtained of a relatively few very nearby galaxies obtained using amateur setups.\footnote{Most investigations to date using amateur telescopes have been biased toward imaging ``pretty'' galaxies with known peculiarities, using  imaging setups without standard filter sets, or even without any filters at all. In many cases this is because the equipment being used is optimized for aesthetic or artistic imaging. Actually, judged as works of art, we think the absolutely spectacular images generated by many of these amateurs far out-distances the images produced by most professional astronomers.} However, ``picking'' between small and large telescopes in this way, where both are limited to similar surface brightnesses, is not nearly as interesting as trying to figure out {\em why} both are constrained to the same limiting surface brightness. And it would be far more interesting still to use this information to find a way to break through the limit, in order to study structures that neither can easily detect.

We now return to consideration of the systematic errors that make low surface brightness imaging very hard.  Most of the obvious systematics\index{systematic errors} are instrumental, and these find their origin in the optical train (e.g., in scattering and internal reflections, which \citealt{slater:09} show quite clearly is the most severe limitation in the case of the Burrell-Schmidt, the most optimized low surface brightness imager on the planet prior to the system described in Sect.~4) or in the detector (e.g., in imperfect flat fielding and dark-current subtraction). Once these have been mastered, nature provides a host of complications external to the imaging system. Some of these complications are well known, such as variability in the sky background introduced by  airglow lines in the upper atmosphere, and some are not so well known, such as the non-negligible structure of the telescopic PSF on spatial scales of tens of arcminutes 
(\citealt{king:71, racine:96, sandin:14}). One must also reckon with very significant sources of low surface brightness contamination that have an extraterrestrial origin, such as Galactic cirrus (likely to be the dominant systematic in many cases) and the unresolved sources making up the extragalactic background light (\citealt{bernstein:07}). 

The reader will probably agree that this is a distressingly long list of systematics\index{systematic errors} that one needs to worry about. In most cases, such as in trying to detect galactic stellar haloes, imperfect corrections made for the systematics just noted will result in {\em false positive} detections. For example, it is not  hard to undertake deep imaging observations and ``find'' galactic stellar haloes that in reality may or may not exist (\citealt{sandin:14,sandin:15}). {\em The true challenge is to hit the required depth with the precision needed to find low surface brightness structures when they really do exist}. 
But in spite of (or perhaps because of) these issues, the low surface brightness Universe is a treasure trove of almost totally unexplored astrophysical phenomena: galactic stellar haloes are only the beginning. If one could ``crack'' the systematics preventing us from getting down to really low surface brightness levels, new avenues would be opened up in a large range of subjects.
A host of phenomena are known to exist at very low surface brightness, ranging from dust rings around planets in our Solar system (\citealt{hamilton:15}), to supernova light echoes (\citealt{rest:12}), to high velocity clouds raining onto the disk of the Milky Way (\citealt{simon:06}), to intra-group and intra-cluster light (\citealt{montes:14}), and even to new emission mechanisms in rich clusters of galaxies (\citealt{yamazaki:15}).
The challenge is to come up with a telescope design that allows the systematic errors just noted to be addressed.

\section{Small Telescope Arrays as Better Imaging Mousetraps}

General-purpose telescopes have to make certain compromises, some of which place limitations on their effectiveness for low-surface brightness imaging. In this Section we are going to examine some of these compromises at a high level, and also describe an alternative telescope concept that is in some ways better-optimized for wide-field imaging than conventional telescope designs. This discussion will continue at a lower level in the next Section by presenting the Dragonfly Telephoto Array\index{Dragonfly telephoto array}, which is an existence proof for the concept. These two Sections may be considered something of a digression, and readers who are more interested in science results than in telescopes may wish to skip ahead to Sect.~\ref{sec:lowsbuniverse}.

Telescopes are designed on the basis of complicated trades between aperture, field of view, resolution, wavelength coverage, and many other factors. In general, the right telescope for the job depends on the job. For a seeing-limited survey of point sources, an appropriate figure of merit\index{figure of merit} $\Phi$ for a telescope is:
\begin{equation} 
\Phi \propto {{D^2 \Omega\, \eta}\over {d\Omega}},
\label{eqn:meritLSST}
\end{equation}
where $D$ is the aperture, $\Omega$ is the field of view, $\eta$ is the throughput and $d\Omega$ is the resolution. Optimizing for this figure of merit drives one toward a telescope design with a fairly large aperture and a large field of view, such as the innovative optical design chosen for the Large Synoptic Survey Telescope (LSST). On the other hand,  if one is more interested in investigating individual targets over a small area of the sky at diffraction-limited resolution, a more appropriate figure of merit is: 
\begin{equation}
\Phi \propto {{D^4 \Omega\, \eta} \over \lambda},
\label{eqn:meritELT}
\end{equation}
\noindent where $\lambda$ is the wavelength. This figure of merit drives the design of telescopes such as the European Extremely Large Telescope (E-ELT) and the Thirty Meter Telescope (TMT), in which both adaptive optics and large aperture play important roles. 

Telescopes with designs optimized on the basis of Eqs.~(\ref{eqn:meritLSST}) and (\ref{eqn:meritELT}) are on the horizon, and these will provide the next generation of astronomers with a truly formidable set of capabilities.  Nevertheless, there are some interesting edge cases that benefit from being looked at in a somewhat different way. For example, consider the situation where one is only interested in imaging very faint structures that are much larger than the limiting resolution of the instrument (or the atmosphere).\footnote{Obviously, for most problems in astronomy, one {\em does} care about resolution, and on the rare occasions when one doesn't, with digital detectors one can always rebin data. That said, we recently scratched our heads at the wisdom of having to do a 15x15 re-binning of beautiful 0.5\,arcsec seeing CFHT and Gemini imaging data in order to obtain signal-to-noise levels comparable to those obtained using telephoto lenses for a number of very diffuse low surface brightness structures.} The figure of merit simply scales with the number of detected photons, so for a very extended source, this will be given by\index{figure of merit}:
\begin{equation}
\Phi \propto \mu\, a\, \eta f^{-2},
\label{eqn:merit}
\end{equation}
where $\mu$ is the surface brightness, $a$ is the pixel area and $f$ is the focal ratio (also known as the $f$-ratio). This equation is interesting because the aperture of the telescope does not enter into the equation directly, but the focal ratio does. Therefore, for imaging of very extended sources, one doesn't necessarily want a big telescope. Instead, one wants an {\em optically fast} (low $f$-ratio) telescope.

Another perspective on Eqs.~(\ref{eqn:meritLSST}) and (\ref{eqn:meritELT}) also depends on considerations of spatial resolution. The $D^4$ scaling in Eq.~(\ref{eqn:meritELT}) refers to cases of diffraction-limited imaging, where the resolution scales as $\lambda/D$. The $D^2$ scaling of Eq.~(\ref{eqn:meritLSST}) refers to cases where the telescope is limited by atmospheric resolution, which is around 0.5\,arcsec at the best sites. This resolution corresponds to the diffraction limit of a 25\,cm (10\,inch) telescope working at visible wavelengths. Therefore, all ground-based telescopes with images that are not sharpened by adaptive optics are operating with the resolution of (at best) a 25\,cm telescope. In that case there is no {\em fundamental} difference between obtaining an image with a large aperture telescope and stacking the images (obtained at the same time) from an array of smaller aperture telescopes. Whether or not there is a {\em practical} difference depends on a myriad number of factors, such as the read noise and dark current in the detectors relative to the poisson noise from the sky background. But, at least in principle, the stacked image is  equivalent to that obtained from a ground-based telescope with aperture $D_{\rm eff}$ and focal ratio $f_{\rm eff}$:
\begin{eqnarray}
D_{\rm eff} & =\sqrt{N} \times D\\
f_{\rm eff} & = f/\sqrt{N},
\end{eqnarray}
\noindent where $N$ is the number of small telescopes, each of which has aperture $D$.

Arrays of small telescopes have a number of advantages and disadvantages\index{advantages and disadvantages of telescope arrays}. One disadvantage is that the need for replication of $N$ instruments to go on the $N$ telescopes, so instruments that are simple to mass-produce (such as cameras) are probably the most suitable ones to deploy on the array. Another disadvantage is  computational complexity, though the last few decades have seen innovations in computation far out-stripping innovations in classical optics, so relying on Moore's law and the inherent parallelism of image stacking means this disadvantage is no longer very serious. Perhaps the most serious disadvantage is mechanical complexity, since $N$ telescopes means there are $N$ more things that can go wrong, presenting a management challenge. The best defence against this disadvantage is more parallelism, because if $N$ is large and all telescopes are completely independent then it does not much matter if a few telescopes are non-operational on a given night. On the other hand, the advantages of this design can be quite remarkable. 
\begin{itemize}
\item The performance of the system can be excellent, because individual telescopes in the array can be relatively simple all-refractive designs that offer superb performance in terms of scattering and ghosting (see next Section). And by making $N$ large enough, systems with $f_{\rm eff} < 0.5$ are possible, which is not possible using monolithic telescopes (since $f=0.5$ is the thermodynamic limit). 
\item Individual telescopes and arrays are inexpensive because they can take advantage of the commercial availability of the needed components. For example, conventional large telescopes have long focal lengths and thus require sensors with large pixels (typically $\gtrsim15\,\mu$m) for which there is essentially no commercial demand. Small telescopes have short focal lengths that are excellent matches to cell phone sensors (whose pixels are typically $\sim1.5\,\mu$m in size). Exponential growth in the demand for such sensors has driven their development to the point where a few dollars can easily buy an off-the-shelf 20\,megapixel sensor whose performance is quite comparable to that of a CCD sensor for a large telescope, but at a cost that is about four orders of magnitude lower.
\item Arrays can start small and grow large relatively inexpensively. For telescope arrays, the cost scales with aperture as $D_{\rm eff}^2$ for both the telescope and the enclosure, which is a much more favourable scaling relation than is the case for conventional large telescopes. The cost of conventional large telescope designs scales with aperture as $D^{2.85}$ (for pre-1980 designs; \citealt{meinel:78}) and $D^{2.5}$ (for post-1980 designs; \citealt{vanbelle:04}). Furthermore, the cost of an enclosure for a conventional large telescope scales as $D^3$ (\citealt{vanbelle:04}).
\end{itemize}


\section{The Dragonfly Telephoto Array}

\begin{figure}[tbh]
\floatbox[{\capbeside\thisfloatsetup{capbesideposition={left,top},capbesidewidth=6cm}}]{figure}[\FBwidth]
{\vspace*{0.3cm}\includegraphics[width=2.1in]{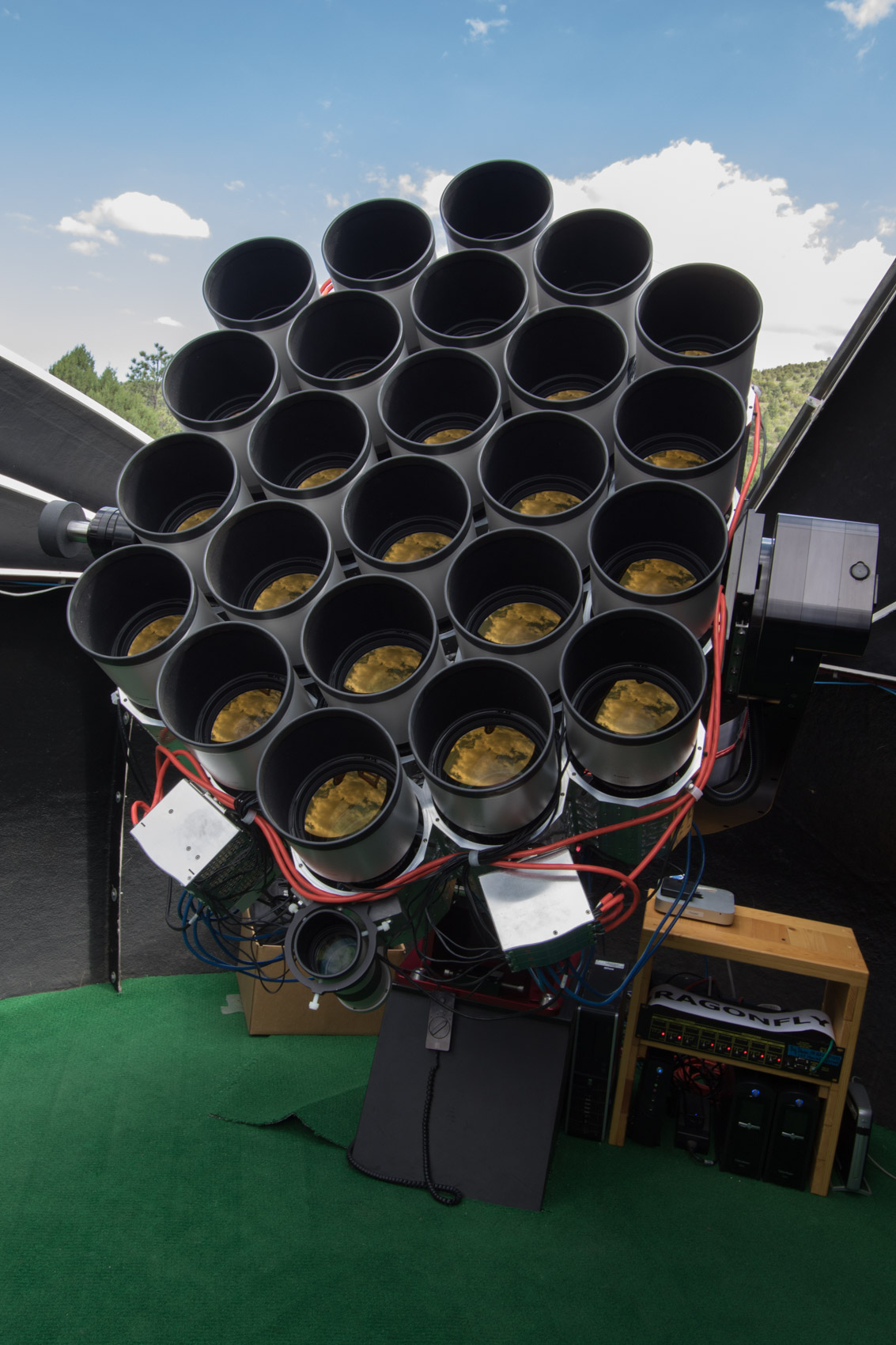}}
{\caption{One of the two 24-lens arrays comprising the 48-lens Dragonfly Telephoto Array. The lenses are co-aligned and the full array is equivalent to a 1m aperture $f/0.39$ refractor with a $2\times 3$\,deg field of view. The arrays are housed in domes at the New Mexico Skies observatory, and operate as a single telescope slaved to the same robotic control system. The lenses are commercial 400\,mm Canon USM IS~II lenses that have superb (essentially diffraction limited) optical quality. This particular lens has very low scatter on account of proprietary nanostructure coatings on key optical surfaces (see Abraham \& van Dokkum 2014 for details). Each lens is affixed to a separate CCD camera and both are controlled by a miniature computer attached to the back of each camera that runs bespoke camera and lens control software that we have made publicly available on a GitHub repository. In the latest incarnation of Dragonfly, each lens is self-configuring and is controlled by its own node.js JavaScript server  in an ``Internet of Things'' configuration that provides a RESTful interface. Growing the array is done by simply bolting a new lens onto the array and plugging in network and power cables
}}
\label{fig1}
\end{figure}

The Dragonfly Telephoto Array\index{Dragonfly telephoto array} (\citealt{abraham:14}) was designed to use the ideas in Sect.~3 in order to break through most of the systematic errors noted in Sect.~2, so we could explore the low surface brightness Universe with high precision. The most intractable problem in low surface brightness imaging is scattering in the optical train, typically from faint stars in the field, though sometimes from bright stars outside the field (see \citealt{slater:09} for a beautiful investigation of the sources of scatter). Because this scattering originates in several of the basic design trades that make large telescopes possible (e.g., an obstructed pupil and reflective surfaces that have high-frequency micro-roughness that backscatters into the optical path and pollutes the focal plane), Dragonfly has an unobstructed pupil and no reflective surfaces at all. The array builds up its effective aperture by multiplexing the latest generation of high-end commercial telephoto lenses that use nano-fabricated coatings with sub-wavelength structures to yield a factor of ten improvement in wide-angle scattered light relative to other astronomical telescopes (see \citealt{sandin:15} for a comparison).  The array is designed to increase in aperture with time, and over the last two years a 48-lens Dragonfly array has been assembled gradually in New Mexico as a collaboration between the University of Toronto, Yale and Harvard. In its current configuration (half of which is shown in Fig.~1) Dragonfly is  equivalent to a 1\,m aperture $f/0.39$ refracting telescope with a six square degree field of view and optical scattering an order of magnitude lower than conventional telescopes. The current detectors have $5.4\,\mu$m pixels resulting in an angular resolution of 2.85\,arcsec/pixel, so the images are under sampled. This is fine for our science goals, and the array has superb performance when imaging low surface brightness structures on scales larger than about 5\,arcsec. However, we can envision a number of programs that would benefit from improved point source sensitivity, and we are planning a detector switch in the future to improve both resolution and point source limiting depth. We are also planning to add a narrow-band imaging mode, the motivation for which will be described in Sect.~5.3.

\begin{figure}[tbh]
\begin{center}
{\vspace*{0.2cm}\includegraphics[width=4.65in]{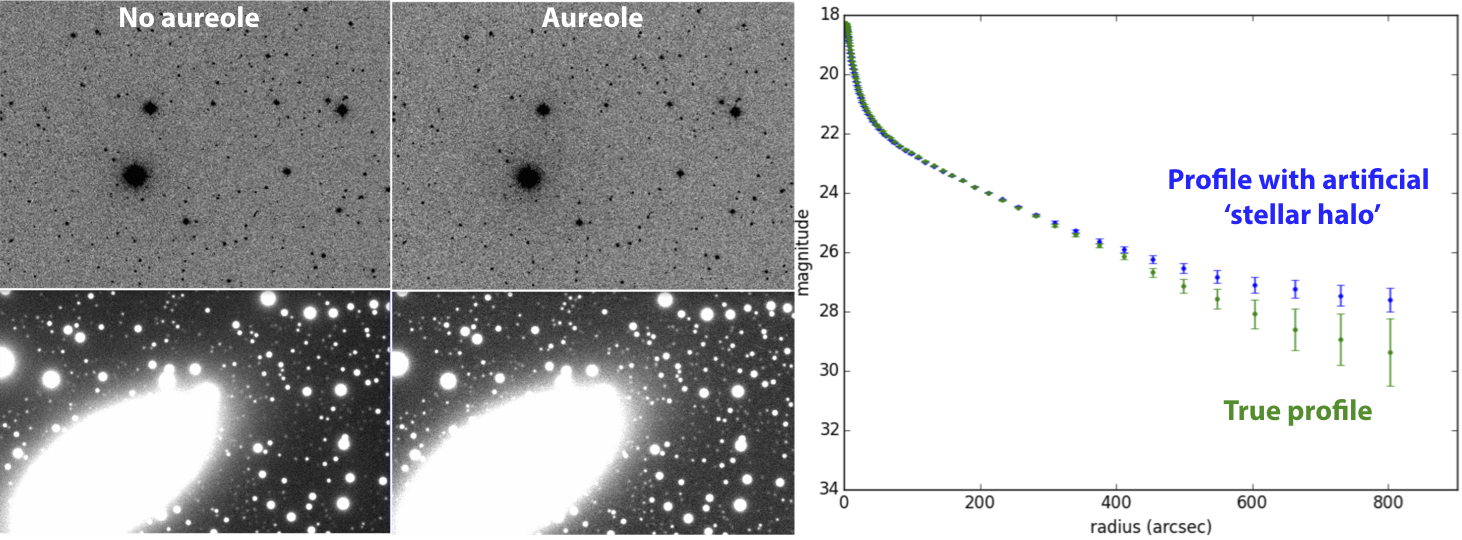}} 
{\caption{
A ``spot the difference'' exercise for the reader. This figure shows the results from a simulation intended to explore the impact of variability on the atmospheric component of the wide-angle stellar PSF (the ``stellar aureole''). The {\it top} panels in the first two columns show $40 \times 60$\,arcmin regions from single 600\,s Dragonfly frames with photometric zeropoints differing by 0.1\,mag. The intensity of the stellar aureole has been found to correlate with the zeropoint, and investigation of fields with bright stars shows that small variations in the photometric zeropoints of individual frames at this level correspond to fields with PSFs that have significantly different structure in their wide-angle wings. The {\it bottom} half of the left two columns show cutouts from simulated observations of a typical target galaxy. The simulation models the results of a $\sim2$\,hour exposure with Dragonfly, made by stacking 600 frames obtained from 48 lenses on a galaxy with the structural properties of NGC~2841. The images adopt different forms for the wide-angle PSF, with shapes that result in a 0.1\,mag difference in the recovered zeropoints. In the {\it left} image we have assumed a standard PSF model out to $\sim 1$\,arcmin with negligible contribution beyond this scale. The {\it right} image shows the corresponding simulation with a stellar aureole included. The prescription for the aureole is taken from Racine (1996). Note that the impact of the aureole is nearly invisible to the eye, except for the merest hint of additional low level scatter. Nevertheless, the aureole has a profound impact on the shape of the surface brightness profile at large radius. The {\it right-hand} panel presents the azimuthally-averaged profile of the target galaxy out to a radius of nearly 0.25\,degree. Because the impact of short-timescale wide-angle PSF variability is significant, in the Dragonfly pipeline individual frames are automatically calibrated and assessed for image quality as they arrive.  In practice around 20\% of data frames that appear fine to the eye are dropped from the final stack because of small variations in their photometric zeropoints. See Zhang et al. (in preparation) for details (reproduced with permission)
}}
\label{aureole}
\end{center}
\end{figure}

Several of the sources of systematic error in ultra-deep imaging find their origins in the atmosphere, rather than in the instrument. To combat these, the operational model for using Dragonfly is in some ways as innovative as the hardware\index{point spread function, wide angle}. When investigating galaxy haloes, the array points only at locations pre-determined (on the basis of IRAS imaging) to have low Galactic cirrus contamination, and the array operates in a fully autonomous robotic mode that tracks atmospheric systematics in real time. The latter point is important because once one has greatly reduced the wide-angle scatter inherent to the instrument, the tall pole becomes the atmosphere. The wide-angle telescopic PSF (the ``stellar auroeole'') is not well understood and at least some of its origin is instrumental (e.g., Bernstein 2007). But a very significant fraction is due to scattering by icy aerosols in the upper atmosphere (a fact well known to atmospheric physicists; see \citealt{devore:13} for an interesting application of stellar aureole measurements to global warming). This component of the wide-angle scatter is variable (a fact noted by Sandin 2014), and our Dragonfly data shows quite clearly that this variability extends down to a timescale of minutes\index{point spread function, variability of}. Dealing with these sorts of ``second order'' non-instrumental scattering issues clearly matters---see Fig.~3 for an illustrative example. We note that some authors correct for wide-angle PSF contamination by post-facto measurement and subsequent modelling (e.g., \citealt{trujillo:2016}), and this is certainly a good thing. However, until the sources of the variability in the atmospheric component of the wide-angle PSF are better understood, we think that real-time monitoring of the large-angle PSF is likely to be important for reliable measurement of stellar haloes. 

\section{The Universe Below 30\,mag/arcsec$^{2}$}
\label{sec:lowsbuniverse}

\subsection{Galactic Outskirts}

Dragonfly has recently completed a campaign of ultra-deep imaging of 18 nearby galaxies (the Dragonfly Nearby Galaxies Survey; \citealt{merritt:16})\index{Dragonfly Nearby Galaxies Survey}. All results from this campaign were obtained with Dragonfly in its prototype eight- and ten-lens configurations, where its performance was equivalent to that of a 0.4\,m $f/1.0$ refractor, and a 0.45\,m $f/0.9$ refractor, respectively. An analysis of the (essentially absent) stellar halo of M101 using Dragonfly data was presented in \citet{vandokkum:14}, and an analysis of the satellite content of M101 appeared in \citet{merritt:14}. \citet{vandokkum:15a}, \citet{vandokkum:15b} and \citet{vandokkum:16} show Dragonfly results on ultra-diffuse galaxies in the Coma cluster (described more in the next Section). Our main focus in this Section will be to highlight results that have recently appeared in \citet{merritt:16}, which presents data on the stellar halo mass fractions of the first eight galaxies in the Dragonfly Nearby Galaxies Survey. We will also highlight preliminary Dragonfly results on the outer disks of galaxies that will be appearing in Zhang et al. (2017) and Lokhorst et al. (2017). 

\begin{figure}[tbh]
\begin{center}
 \includegraphics[width=4.7in]{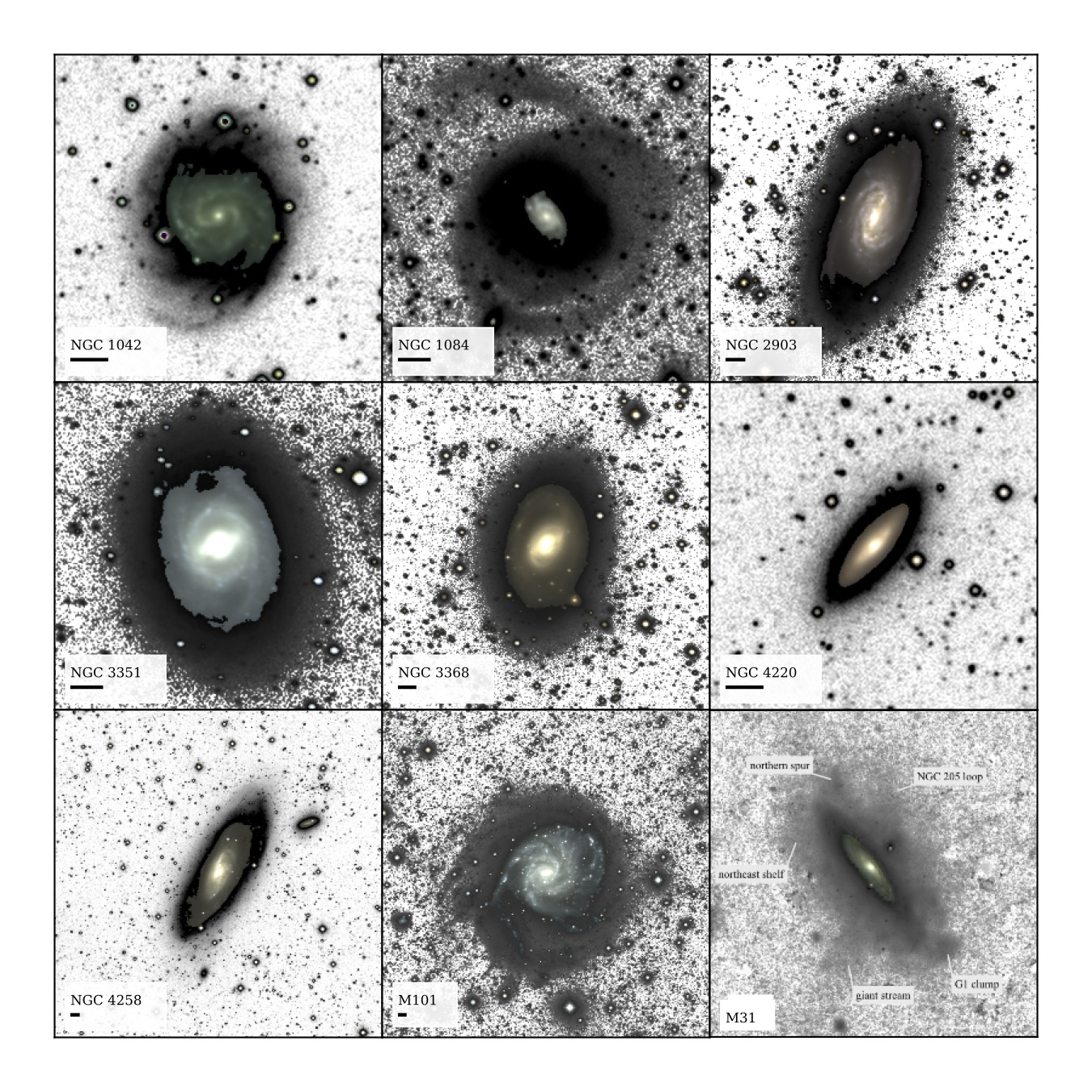} 
 \caption{Images of each of the eight galaxies in the sample of \citet{merritt:16}. The
    pseudo-colour images were created from $g$ and $r$-band images for
    the high surface brightness regions, and the greyscale shows the
    lower surface brightness outskirts. The {\it bottom-right} panel shows
    M31, created from a combination of Dragonfly and PAndAS
    data (\citealt{mcconnachie:09}, \citealt{carlberg:09}) and redshifted to a
    distance of 7 \,Mpc (\citealt{vandokkum:14}). Black lines beneath each galaxy
    name indicate scales of 1 arcmin. Figure taken from \citet{merritt:16}, reproduced
    with permission
    \label{allspirals}}
\end{center}
\end{figure}

The data shown here are taken from the Dragonfly Nearby Galaxies Survey (\citealt{merritt:16}), which is a sparse-selected ultra-deep imaging study of 18 galaxies selected on the basis of four criteria: (1) galaxies must have $M_B < -19$\,mag; (2) galaxies must be further than 3 \,Mpc away; (3) galaxies must be located at high Galactic latitude within ``holes'' of low Galactic cirrus determined from IRAS 100$\,\mu$m imaging; (4) galaxies must be visible for extended periods of time with low air mass as seen from New Mexico.  No other selection criteria were imposed. The third criterion is interesting, because it turns out that most regions of the high Galactic latitude sky are unsuitable for ultra-deep imaging without pre-selection to avoid cirrus contamination\index{stellar halo}.

Deep images of the first eight galaxies (all spirals) in the Dragonfly Nearby Galaxies Survey are shown in Fig.~\ref{allspirals}, together with a ``reconstructed'' (by combining star-count data in the outskirts and Dragonfly data in the interior) image of M31 as seen at a distance of 7 \,Mpc using the same instrument. Each image is $30$ arcmin on a side. The familiar high surface brightness appearance of the galaxies is shown in colour; the low surface brightness outskirts are shown using a greyscale. The galaxies show a remarkable diversity in their low surface brightness structures. Nevertheless, it seems that M31-like haloes, characterized by significant substructure, do not appear to be the norm. 

\begin{figure}[tbh]
\begin{center}
 \includegraphics[width=4.7in]{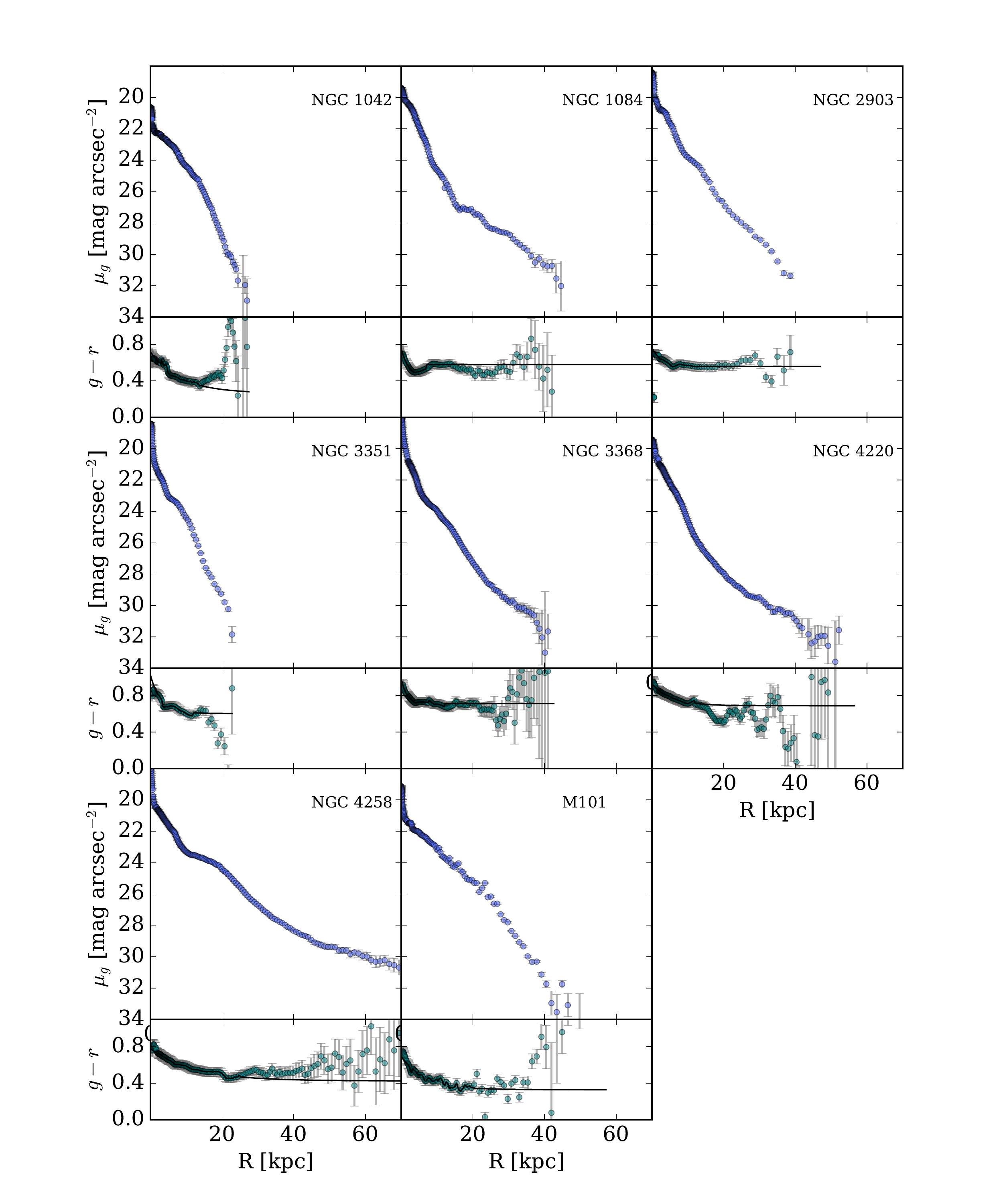} 
\caption{
    Surface brightness and $g-r$ colour profiles (with
    $1 \sigma$ error bars) for the first eight galaxies explored in 
    the Dragonfly Nearby Galaxies Survey. Figure taken from Merritt et al. (2016),
    reproduced with permission
    \label{egmug}}
\end{center}
\end{figure}

Surface brightness profiles\index{surface brightness profiles} corresponding to the galaxies shown in Fig.~\ref{allspirals} are shown in Fig.~\ref{egmug}. All profiles extend down to at least 32\,mag/arcsec$^2$, with some stretching down to 34\,mag/arcsec$^2$. We have restricted these profiles to regions with high signal-to-noise, and have incorporated known systematics into the uncertainty estimates. Since all of the galaxies are well-studied, Merritt et al. (2016) presents careful comparisons between profiles obtained using Dragonfly data and previously published surface brightness profiles for these galaxies. There is an excellent agreement between these profiles in brighter regions where many surveys overlap. Results remain quite consistent until around 29\,mag/arcsec$^2$, at which point relatively little comparison data exists. However, starting at these low surface brightness levels, Dragonfly profiles generally show less evidence for deviations from exponential disks than do profiles from other investigations (e.g., \citealt{pohlen:06} and \citealt{watkins:14}). One exception is NGC~4258, since the Dragonfly observations of this system reveal a very low surface brightness extended red structure. The reader is referred to Merritt et al. (2016) for details.

What can one learn from these profiles? Probably their most remarkable characteristic is the fact that only a few galaxies show evidence for prominent upturns at low surface brightness levels that might signal the presence of a stellar halo built up by coalescing substructures. If they existed, substructures from M31-like haloes would be expected to dominate the profiles below about 27.5\,mag/arcsec$^2$ (\citealt{bakos:12}), while the very faint streams predicted by numerical simulations (e.g., \citealt{johnston:08}) would likely dominate profiles below around 30\,mag/arcsec$^2$. While some objects such as NGC~1084 (the existence of substructure in which was already known from \citealt{martinez-delgado:10}) do show substructures at a level reminiscent of M31, three objects (M101, NGC~1042, and NGC~3351) show no evidence at all for an extended stellar halo---these profiles appear to be dominated by flux from the disk to the limits of our observations. 

\begin{figure}[tbh]
\begin{center}
 \includegraphics[width=4.6in]{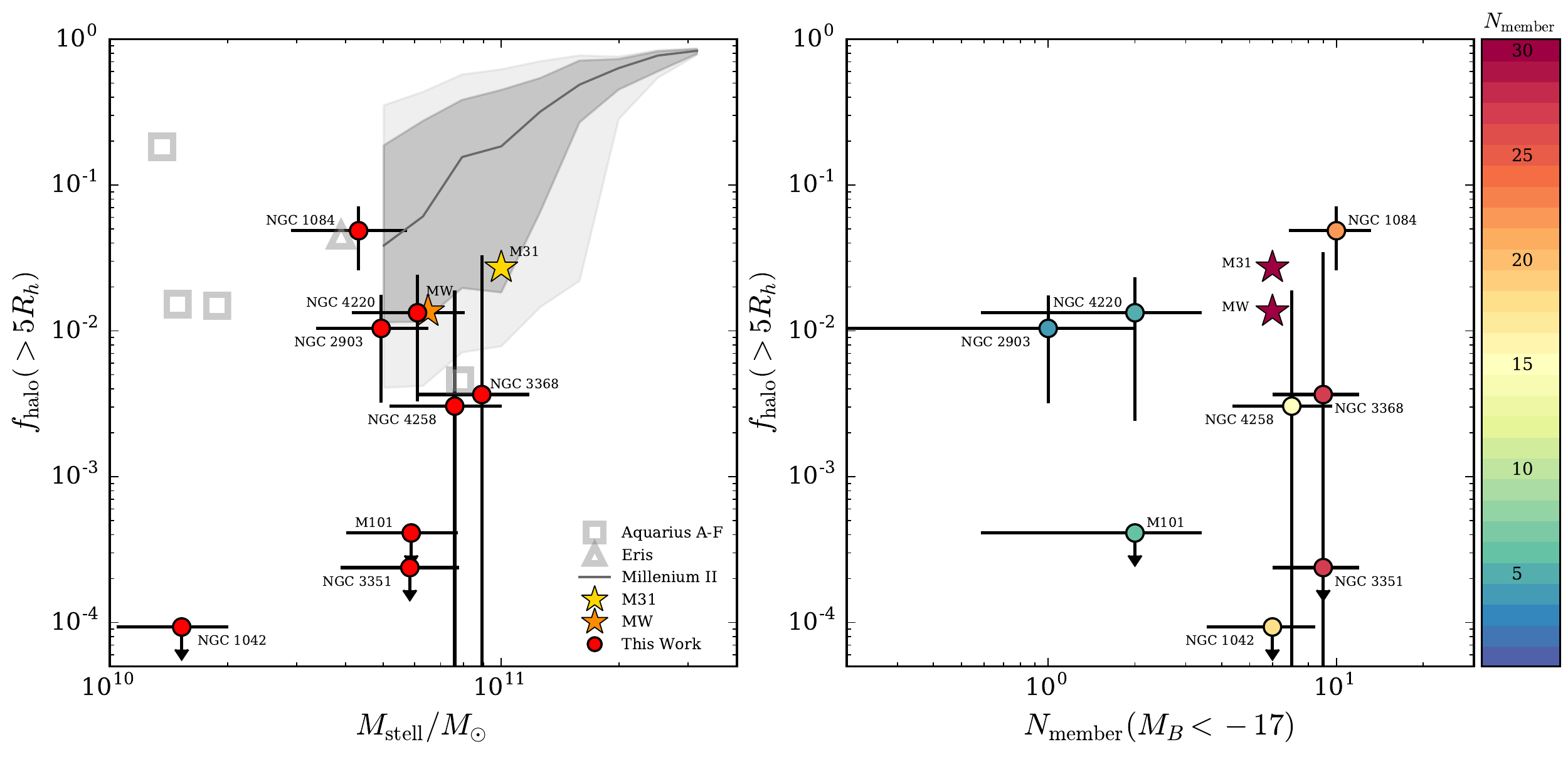} 
\caption{{\it Left}: The stellar halo mass fractions (and $1 \sigma$ errors) for the sample in Merritt et al. (2016), measured beyond $5\,R_{\rm h}$ (red points). Values of $f_{\rm halo}$ for the Milky Way (\citealt{carollo:10}) and M31 (\citealt{courteau:11}) are shown for comparison (orange and gold stars, respectively), and have been scaled to the halo mass fraction outside of $5\,R_{\rm h}$ assuming the structure of the halo of M31 (\citealt{irwin:05, courteau:11}). Predictions of $f_{\rm halo}$, measured over $3 \leq r \leq 280$\,kpc from the Aquarius simulations (\citealt{cooper:10}); over $r \geq 20$\,kpc from the Eris simulation (\citealt{pillepich:15}); and over $r \geq 3$\,kpc from the Millennium II simulation (galaxies with $B/T < 0.9$ only; \citealt{cooper:13}) are indicated by grey open squares, triangles, and shaded region, respectively. {\it Right}: Environmental richness is parametrized by the number of group members (\citealt{makarov:11}) with $M_{B} < -17$. The colour of each symbol corresponds to the \textit{total} number of known group members for that particular galaxy. The stellar halo mass fractions do not appear to be a function of environment. Figure taken from Merritt et al. (2016), reproduced with permission.
\label{fhalo}}
\end{center}
\end{figure}

A complementary approach to understanding the contribution from stellar haloes is presented in Fig.~\ref{fhalo}, which shows the stellar halo mass fraction as a function of total stellar mass\index{stellar halo mass fraction}. The total stellar masses of nearby haloes are largely unexplored (with some notable exceptions, such as \citealt{seth:07, bailin:11, greggio:14, vandokkum:14, streich:15}). For concreteness, we define the halo mass fraction to be the stellar mass in excess of a disk$+$bulge model outside of five half-light radii $R_h$, a region where the stellar halo should start to contribute significantly (\citealt{abadi:06, johnston:08, font:11, cooper:15, pillepich:15}). Given the relatively narrow range in stellar mass explored by these observations ($2-8\times 10^{10}M_{\odot}$), the data display a remarkably wide range in stellar halo mass fractions. One of the galaxies in our sample, NGC~1084, has a stellar halo mass fraction of $0.049 \pm 0.02$ (even larger than that of M31), while, as noted earlier, three others (NGC~1042, NGC~3351, and M101) have stellar haloes that are undetected in our data. We measure an RMS scatter of $1.01^{+0.09}_{-0.26}$ dex, and a peak-to-peak span of a factor of $>100$. This level of stochasticity is high and certainly exceeds the expectations (\citealt{amorisco:15, cooper:10, cooper:13}) from numerical simulations (the grey regions shown in the Figure), though they may be qualitatively consistent with variations in the structure and stellar populations of nearby stellar haloes observed in both integrated light and star counts studies (e.g., \citealt{mouhcine:07, barker:12, monachesi:15}). The former may be contaminated by scattered light, and the latter (based on pencil beams) may not be fairly sampling the haloes, so the Dragonfly observations provide a robust baseline for future characterization of the stellar halo mass fraction in luminous nearby spirals.

\begin{figure}[tbh]
\begin{center}
 \includegraphics[width=4.6in]{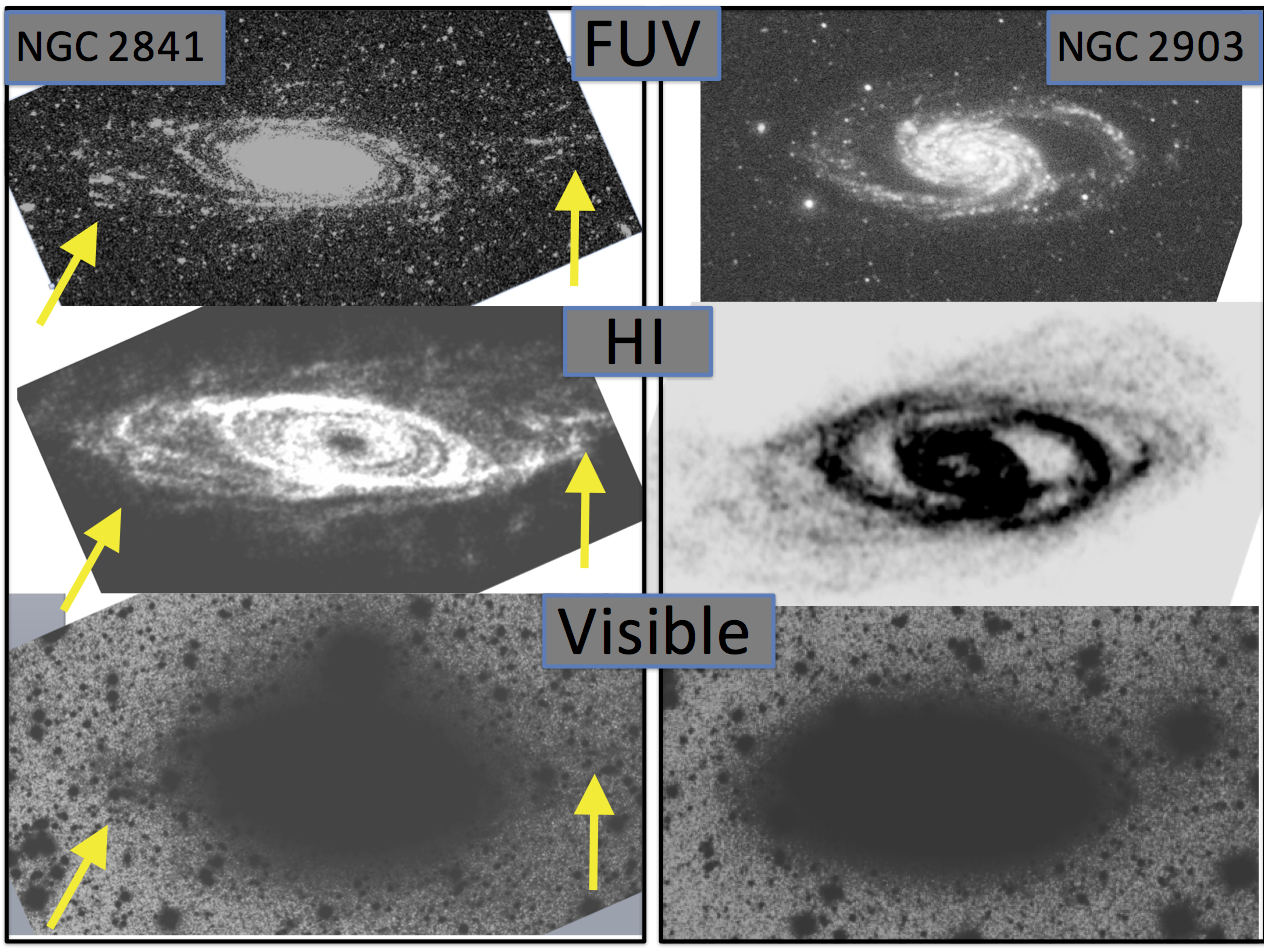} 
 \caption{{\it GALEX} FUV images ({\it top row}), H{\sc i} observations from THINGS ({\it middle row}), and ultra-deep Dragonfly imaging ({\it bottom row}) for NGC~2841 ({\it left}) and NGC~2903 ({\it right}), two systems showing extended UV disks (\citealt{thilker:07}). Taken from Zhang et al. 2017 (in preparation), reproduced with permission}
   \label{fig:pandragon}
\end{center}
\end{figure}

\begin{figure}[tbh]
\begin{center}
 \includegraphics[width=4.6in]{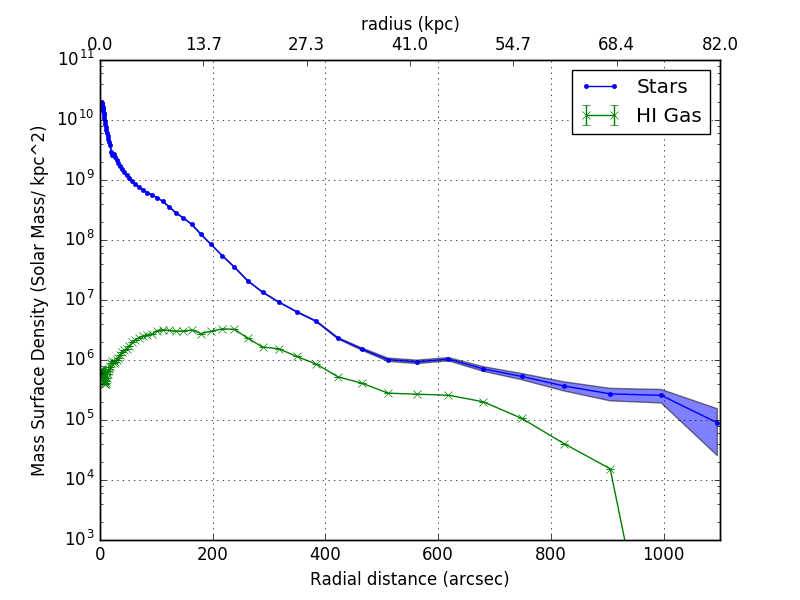} 
 \caption{The stellar mass surface density profile ({\it top}) and neutral gas mass density profile ({\it bottom}) of NGC~2841 based on the Dragonfly observations shown in the previous figure. Taken from Zhang et al. 2017 (in preparation), reproduced with permission}
   \label{fig:stargasprof}
\end{center}
\end{figure}

One possible criticism of the results presented so far is that they are based on analyses of surface brightness profiles\index{extended stellar disks}. The obvious benefit of using profiles is the increase in signal-to-noise they bring because of averaging along isophotes. Another very useful benefit of profiles is that they decrease the dimensionality of the problem being investigated. However, a potential weaknesses of profiles is their strong sensitivity at large radii to the accuracy of sky background estimates. At a more fundamental level, profiles might also be criticized for making strong assumptions about an underlying symmetry in the images. It is therefore interesting to note that many of the most important conclusions obtained so far can be reinforced, at least at a qualitative level, simply by a careful inspection of the images. This is particularly true when the data are viewed in the appropriate panchromatic context. Fig.~\ref{fig:pandragon} (taken from Zhang et al. 2017, in preparation) shows a comparison of {\it GALEX}\index{GALEX} FUV images (top row), radio observations of H{\sc i} from THINGS (middle row), and ultra-deep $g$-band Dragonfly imaging (bottom row) for two galaxies with extended UV disks. Lightly-binned Dragonfly images probe down to around 30\,mag/arcsec$^2$ even without the benefit of radial averaging, and at these surface brightness levels it is clear that {\em at visible wavelengths starlight stretches out to the limits of the HI observations}. In the examples shown, the bulk of the starlight at very large radii is contained within extended disks, rather than in a roughly spherical halo. One suspects that the traditional view that at radio wavelengths disks are about twice as big as they are at visible wavelengths is simply the product of the high sensitivity of the radio data and the limited ability of ground-based telescopes to undertake low surface brightness observations. Careful inspection of Fig.~\ref{fig:pandragon} shows that at large radii the starlight in the Dragonfly images traces the H{\sc i} data closely. Smooth starlight also fills in the regions between the UV knots in the {\it GALEX} images, which (like the evolved starlight) stretch out to the limits of the radio observations. The existence of evolved disk stars at very large radii is another manifestation of the important problem first highlighted by the existence of XUV disks (\citealt{thilker:07}): how does one form stars at radii well beyond the bulk of the molecular gas in galaxies, and at locations where disks are at least globally Toomre stable? Certainly local regions of instability can emerge from dense pockets of gas compressed by turbulence (\citealt{elmegreen:06}), but what drives the turbulence, and where does the molecular gas come from? In any case, it is amazing to see in these data how, at least in some cases, giant disks rather than stellar haloes define the faint outskirts of the galaxies. 

Returning to profiles, it is interesting to explore the relationship between the gaseous and stellar components of the outskirts of galaxies. A comparison between stars and gas for one well-studied galaxy, NGC~2841, is shown in Fig.~\ref{fig:stargasprof}\index{extended gaseous disks}. Between 30 and 50\,kpc the two components appear to track each other rather closely. Beyond 50\,kpc the stellar profile flattens while the gas profile appears to decline, though the significance of this is presently unclear given the rather large systematic errors at these radii. Perhaps the upturn in the stellar mass density profile signals the onset of a well-defined stellar halo in this system? This seems plausible, but perhaps more ambitious imaging would also reveal a further continuation of the enormous stellar disk already uncovered, stretching out to radii at which the disk-halo interface fuels the fire of star formation in the galaxies, and ultimately connects these systems into the cosmic web of primordial gas (Sect.~5.3).  

\subsection{Ultra-Diffuse Galaxies}

\begin{figure}[tbh]
\begin{center}
 \includegraphics[width=4.6in]{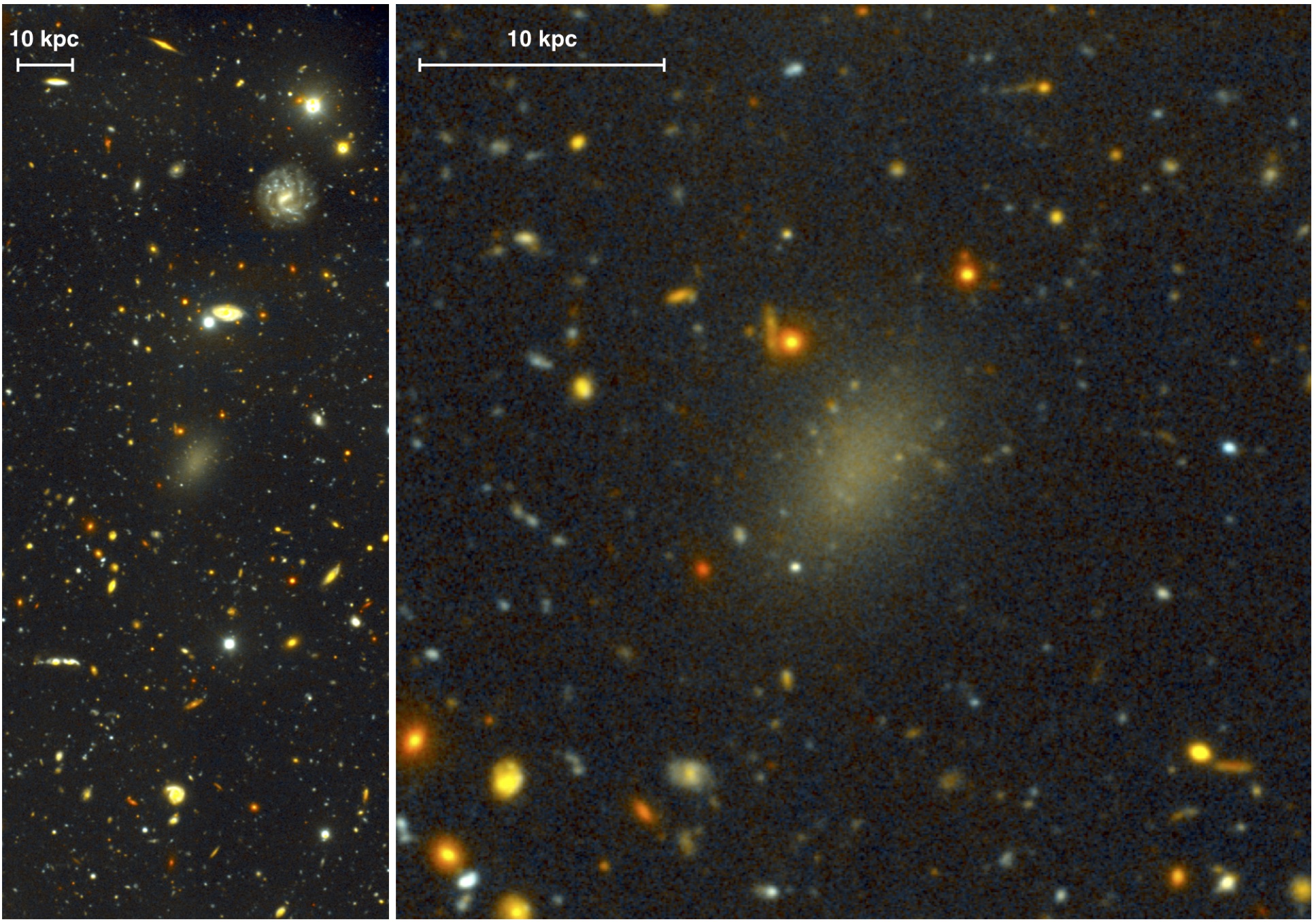} 
 \caption{
Deep Gemini $g$ and $i$ images combined to create a colour image
of the ultra-diffuse Dragonfly~44 and its immediate surroundings. This system
is in the Coma cluster and it has a remarkable appearance:
it is a low surface brightness, spheroidal system whose outskirts are
peppered with faint, compact sources we have identified as
globular clusters. Figure taken from 
\citet{vandokkum:16}, reproduced with permission
 }
   \label{fig:udg}
\end{center}
\end{figure}

One of the first science targets imaged with Dragonfly was the Coma cluster of galaxies\index{Coma cluster}. This is the nearest rich cluster of galaxies and it has been studied extensively for nearly a century, yet Dragonfly's first observation of Coma yielded an interesting discovery: the existence of a large number of very faint, large low surface brightness objects that turned out to be a new class of giant spheroidal galaxies (\citealt{vandokkum:15a, vandokkum:15b}). We named this class of objects {\em ultra-diffuse galaxies} (UDGs)\index{ultra-diffuse galaxies}, because they have sizes similar to the Milky Way (half-light radii around 3\,kpc) but only 1/100 to 1/1000 the number of stars as the Milky Way. Figure~\ref{fig:udg} shows a now well known example, Dragonfly 44\index{Dragonfly 44}, along with its surprisingly abundant globular cluster distribution stretching out into the outskirts of the galaxy. The total number of globular clusters\index{globular clusters around UDGs} can be used to trace Dragonfly 44's dynamical mass via the proportionality between globular cluster numbers and halo mass (\citealt{harris:13, harris:15, hudson:14}), backed up by ultra-deep Keck spectroscopy (\citealt{vandokkum:16}). This particular system has a dynamical mass similar to that of the Milky Way.

The discovery of UDGs\index{ultra-diffuse galaxies, models for} has generated tremendous interest in the community, from observers who are rapidly enlarging the UDG samples (e.g., \citealt{koda:15, mihos:15, vanderburg:16}), from simulators who must now try to understand the origin and evolution of these galaxies (e.g., \citealt{yozin:15,amoriscoloeb:16,amorisco:16}), and even from alternative-gravity researchers who claim their existence challenges dark matter models (\citealt{milgrom:15}). The existence of so many presumably ``delicate'' UDGs (Koda et al. put their number at $\sim800$ in Coma) in a rich cluster poses the immediate question of why they are not being ripped apart by the tidal field of the cluster. They may be short-lived and be on their first infall and about to be shredded, but this seems unlikely given their predominantly old and red stellar populations and smooth morphologies. If they have survived for several orbits in the Coma cluster, then simple stability arguments suggest that they must have significantly higher masses than implied by their stellar populations; in fact, in order to survive, their dark matter fractions need to be $> 98\%$ (\citealt{vandokkum:15a}) within their half-light radii.
 
A central goal of future Dragonfly research is to determine the origin of UDGs in rich clusters. It has been suggested that UDGs are low mass galaxies that are anomalously large (``inflated dwarfs'') because they are undergoing tidal disruption (\citealt{collins:14}) or have a high spin (\citealt{amoriscoloeb:16}). As noted earlier, an alternative idea is that they are very massive dark matter haloes that are greatly under-abundant in stars because they did not manage to form normal stellar populations (\citealt{vandokkum:15b,agertz:15,yozin:15,vanderburg:16}). In this picture UDG's are ``failed giant'' galaxies. Distinguishing between inflated dwarfs and failed giants can be done using dynamical information obtained from absorption line widths, though these are biased towards the interior of the systems. As noted earlier, an alternative approach (that is also being pursued with {\it HST}) is to probe the dark matter content of UDGs by imaging their globular cluster distributions\index{ultra-diffuse galaxies}. Studies with {\it HST} (and soon with the {\it James Webb Space Telescope}) will be a marked improvement on ground-based imaging such as that shown in Fig.~\ref{fig:udg}. Intriguingly, \citet{peng:16}  claim that one object with {\it HST} data (Dragonfly 17) represents the most extreme conceivable case of a ``failed giant'', namely a system comprised almost entirely of dark matter in which the visible stars are all the product of the initial formation of a stellar halo that somehow failed to initiate the formation phase of the bulk of the galaxy. In this case, we need not probe the outskirts of the system to explore the halo---the whole galaxy is a halo! We find a similar situation for Dragonfly 44 (van Dokkum et al. 2016), though there are also many systems with lower mass, suggesting UDGs are a mixed set of objects, and a healthy fraction are dwarfs, as noted by \citet{beasley:16a} and \citet{roman:16}. 

\subsection{Imaging the Cosmic Web---the Next Frontier?}

The ultimate limits of the Dragonfly concept are not yet known. Our robotic operational model and tight control of systematics (in particular our real-time modelling of sky variability) allows unusually long integration times to be undertaken. With the 10-lens array (as the 48-lens array has only just come on-line) our longest integrations are 50\,h in duration (spread over many nights), and over this period of time we have not yet become limited by any systematics (e.g., scattering, sky variability, flat field accuracy) that would make longer integrations pointless. It appears that with the current configuration the ultimate systematic limit on depth will probably be source confusion from unresolved background galaxies, but we can get a factor of $2-3$ improvement in resolution with new CMOS detectors using smaller pixels. (The relevant detector technology is driven by demand for better mobile phone camera sensors, so advances happen quickly). As a result, a \mbox{$\sim500$-lens} Dragonfly array (equivalent to a $\sim3$\,m aperture $f/0.13$ refractor) may well be perfectly feasible and we can already envision ways to grow Dragonfly up to that scale. 

What would one do with a 500-lens Dragonfly? Certainly it would be exciting to go much deeper on our targets, and it would be wonderful to be able to do everything in this Chapter $10\times$ faster. It might be even more exciting to embark on new science objectives such as  taking the exploration of galactic outskirts to a completely new level by adding a narrow-band capability to Dragonfly. This would allow detailed exploration of feedback at the disk-halo interface of the circumgalactic environment. In narrow-band imaging mode Dragonfly could be used to study extremely faint nebular emission lines (primarily H$\alpha$ and [O{\sc iii}]) in the circumgalactic medium (CGM) of nearby galaxies to probe the disk-halo interface (\citealt{putman:12}). Maps of gas flows into and out of galaxies (as inflowing material ``fuels the fire'' of star-formation while winds drive gas back into haloes) would be interesting, and would do much to constrain galactic feedback models.

More speculatively, a large array of small telescopes working in narrow-band imaging mode might allow one to directly image ionized gas in the brighter portions of the filamentary cosmic web. The intergalactic medium (IGM), together with its close cousin the CGM\index{circumgalactic medium}, are arguably the most important (since they contain the most baryons) and least understood (since they're relatively difficult to detect) baryonic components of the Universe. Of course the denser pockets of the IGM/CGM have  long been studied in absorption using UV lines (e.g., the Ly$\alpha$ forest)  and in 21\,cm emission using radio telescopes. However, both approaches are quite limited. Absorption line studies probe limited lines of sight. Furthermore, Ly$\alpha$ must be cosmologically band-shifted in order to be accessible using ground-based telescopes, so more is known about the IGM/CGM at high redshifts than is known locally. At much longer wavelengths, single-dish radio telescopes have the required sensitivity to probe 21\,cm emission from H{\sc i} in haloes in the nearby Universe ($z<0.1$), but they lack the needed resolution, while radio interferometers have the required resolution but they lack the necessary dynamic range. Given these limitations, direct observation of local Ly$\alpha$ emission (the primary coolant for the IGM) from structures in the cosmic web would be extraordinarily interesting. The UV is inaccessible from the ground, but direct imaging of Ly$\alpha$  is already one of the core motivations for the proposed French/Chinese {\it MESSIER} satellite. Could something similar be achieved from the ground?

\begin{figure}[tbh]
\begin{center}
 \includegraphics[width=4.6in]{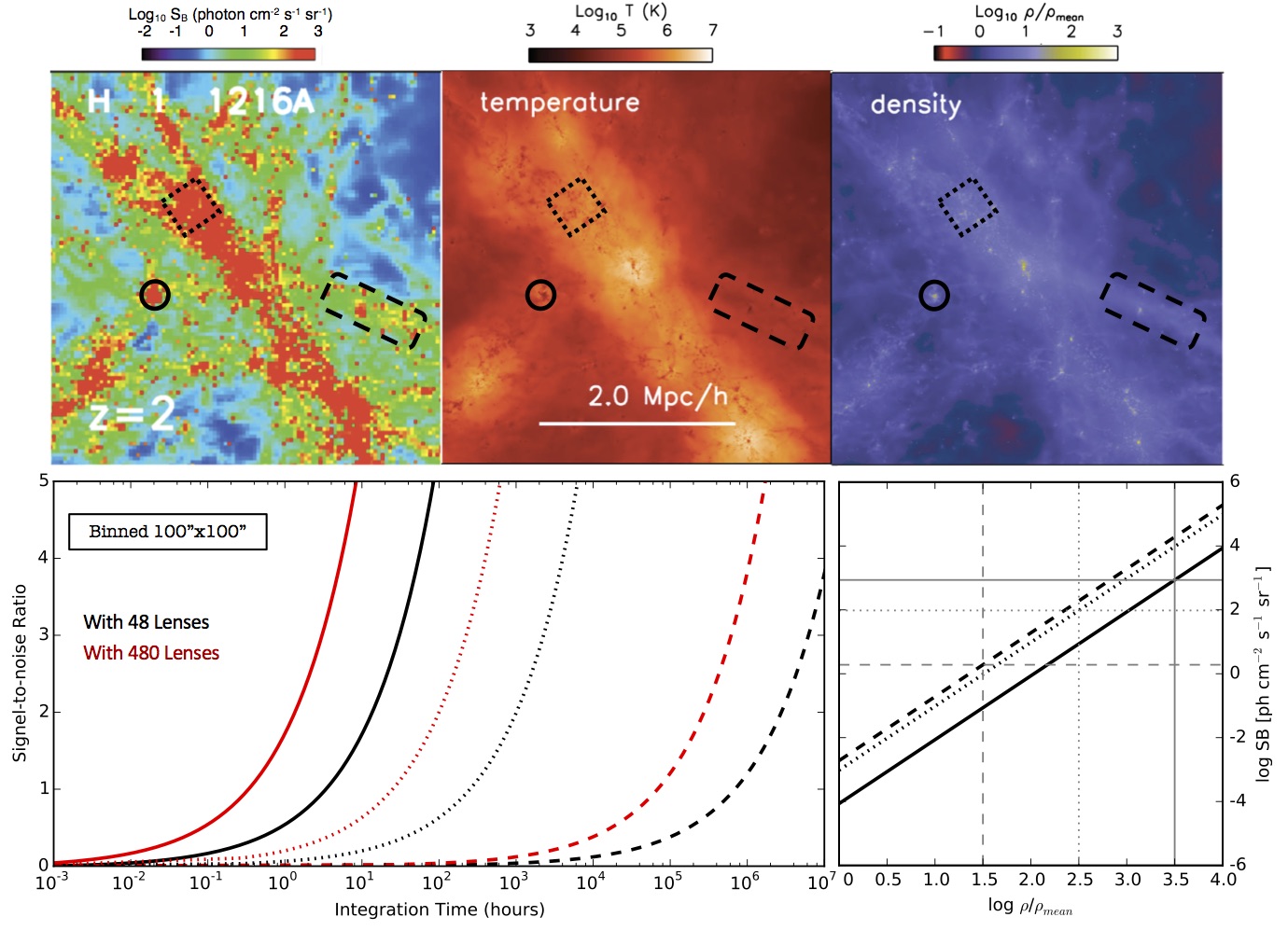} 
\caption{
{\it Top}: The simulated Ly$\alpha$ surface brightness, temperature, and density maps from Figs.~2 and 4 of \citet{bertone:12}, showing the intersection of IGM filaments.  The side of the box is 4 comoving Mpc\,$h^{-1}$. Regions corresponding to $\log_{10}(T) \sim 6$ and $\log_{10}(T) \sim 4-5$ are indicated by dashed boxes and solid circles, respectively.
{\it Bottom left}: The SNR as a function of exposure time for the surface brightness of H$\alpha$ corresponding to the boxed and circled regions in the upper panel. Regions have been binned to $100\times 100$\,arcsec boxes, and red and black lines denote the existing Dragonfly (red) and a system with $10\times$ the current number of lenses. \textit{Bottom right}: The H$\alpha$ surface brightness as a function of the hydrogen density in units of the mean density of the gas. Taken from Lokhorst et al. (2017, in preparation), reproduced with permission
\label{fig:web}}
\end{center}
\end{figure}

Detecting H$\alpha$ emission from the IGM using a ground-based telescope would be a huge challenge, because only $\sim5\%$ of Ly$\alpha$ photons ultimately wind up being re-emitted as H$\alpha$ photons. Simulations (Lokhorst et al. 2017, in preparation) suggest that small telescope arrays could target three aspects of the IGM/CGM: (i) the fluorescent ``skin'' of local ``dark'' H{\sc i} clouds; (ii) extended haloes analogous to the extended Ly$\alpha$ haloes/blobs recently detected around Lyman break galaxies at high redshifts; and (iii) emission from filaments of the IGM itself. Detection of the first two of these would be highly interesting and would likely be achievable with the current 48-lens Dragonfly array using long integration times (tens of hours). The third component poses a much greater challenge. Fig.~\ref{fig:web}, taken from Lokhorst et al. 2017 (in preparation), presents the predicted surface brightness of Ly$\alpha$  from the diffuse gas and dense gas clumps in the Bertone et al. (2012) hydrodynamical simulation of the $z=2$ IGM. We use these simulations to model the corresponding distribution of H$\alpha$ signal-to-noise and surface brightness for various regions in the simulation. Assuming qualitatively similar structures and surface brightnesses in the local Universe, and assuming the required very long integration times can be undertaken without hitting a ``wall'' of systematic errors, direct imaging of dense clumps of the IGM is within the capability of the existing 48-lens Dragonfly imaging in H$\alpha$ when operating in a highly binned low-resolution mode, but investigation of the dense IGM at high resolution, or of more representative emission from the IGM, would require a 500-lens Dragonfly. With such an instrument, one could take the exploration of galactic outskirts to its logical conclusion, by pushing out beyond galactic haloes and into the cosmic web.

\bibliographystyle{spbasic}
\bibliography{abrahamRefs}



\end{document}